\documentclass{aastex}
\usepackage{spr-astr-addons}
\usepackage{url}
\usepackage{natbib}

\RequirePackage{color}

\def\aap{Astron. Astrophys.}

\def\aj{Astrn. J.}
\def\apj{Astrophys. J.}
\def\apjs{Astrophys. J. Suppl.}
\def\apss{Astrophys. Space Science}
\def\apjl{Astrophys. J. Letters}

\def\gca{Geochim. Cosmochim. Acta}
\def\mnras{Mon. Not. R. Astron. Soc.}
\def\nat{Nature}

\def\bain{Bull. Astron. Inst. Netherlands}
\def\aapr{Astron. Astrophys. Rev.}

\def\araa{Annu. Rev. Astron. Astrophys.}
\def\ssr{Space Science Review}

\begin{document}

\title{The feedback of massive stars on interstellar astrochemical processes}
%\slugcomment{Not to appear in Nonlearned J., 45.}
%% Running heads
\shorttitle{Influence of massive stars on astrochemistry}
\shortauthors{De Becker}

\author{M. De Becker}
\affil{Department of Astrophysics, Geophysics \& Oceanography, University of Li\`ege, 17 all\'ee du 6 Ao\^ut, Sart-Tilman 4000, Belgium}

\email{Michael.DeBecker@ulg.ac.be}

\begin{abstract}
Astrochemistry is a discipline that studies physico-chemical processes in astrophysical environments. Such environments are characterized by conditions that are substantially different from those existing in usual chemical laboratories. Models which aim to explain the formation of molecular species in interstellar environments must take into account various factors, including many that are directly, or indirectly related to the populations of massive stars in galaxies. The aim of this paper is to review the influence of massive stars, whatever their evolution stage, on the physico-chemical processes at work in interstellar environments. These influences include the ultraviolet radiation field, the production of high energy particles, the synthesis of radionuclides and the formation of shocks that permeate the interstellar medium.
\end{abstract}

\keywords{Massive stars ; Astrochemistry ; Interstellar medium ; Cosmic rays}

\section{Introduction}
Since the identification of the first interstellar molecule (CH radical) in the 1930's by \citet{ch-swings}, many molecules have been detected in space \citep[see e.g.][and references therein]{ehrenfreund2000,herbst2009,debeckerachem2013}. These discoveries have led to substantial efforts to understand physico-chemical phenomena likely to take place in the interstellar medium. Astrochemistry is the discipline that studies these processes. Considering the particular physical conditions found in interstellar environments, special care must be devoted to unusual factors such as the ultraviolet (UV) radiation field and galactic  cosmic rays (GCR). In addition, the low densities (n\,=\,10$^4$ to 10$^6$\,cm$^{-3}$) and temperatures (T\,=\,10 to 50\,K) in molecular clouds -- with respect to terrestrial standards (n\,$\sim$\,10$^{19}$\,cm$^{-3}$, T\,$\sim$\,300\,K) -- suggest rather slow chemical kinetics. The time scales for many gas phase reactions are comparable to or greater than the evolution time scales of astronomical environments. As a result, one has generally to deal with time-dependent (out of equilibrium) chemistry, and alternative pathways have thus been explored to explain the current census of astromolecules, including more than 170 different species. Surface processes on interstellar grains are essential \citep{grain1998,cazaux,tielens} and dust grains have been determined to act as catalysts, speeding up the kinetics of chemical processes in interstellar clouds.

Stars -- in particular the most massive ones -- play a crucial role in influencing the formation of molecules in space. The usual criterion adopted to consider a star to be massive is its capability to trigger central burning of carbon. This translates into a lower limit on the stellar mass of about 8-10 solar masses \citep{chiosimaeder1986}, with an upper limit located at about 100 solar masses. On the main sequence, i.e. the evolution phase of a star corresponding to central nuclear fusion of hydrogen to yield helium, massive stars are characterized by high effective temperatures (T$_\mathrm{eff}$, typically between 20000 and 50000\,K), and this corresponds to spectral types early-B and O \citep{lanzhubenyO,martins2005,lanzhubenyB}. As a result of their huge luminosity, these stars produce strong stellar winds \citep[see e.g.][]{cak,owocki1990,pulsreview2008,owocki2011} responsible for the input of large amounts of material and mechanical energy into the interstellar medium. Typical mass loss rates are of the order of 10$^{-7}$--10$^{-6}$ solar mass per year, with wind velocities reaching values of the order of 2000--3000\,km\,s$^{-1}$ \citep{muijres2012}. As massive stars evolve, they become Wolf-Rayet (WR) stars or even Luminous Blue Variables \citep{wrcrowther}, characterized by significantly higher (one or two orders of magnitude) mass loss rates as compared to their O and early-B progenitors. Their evolutionary pathway ends either in an explosion as a supernova (SN) or via a Gamma-Ray Burst \citep{georgy2009}. Despite the rarity of these stars ($\sim$ a few ppm of the Galactic stellar population), they play a crucial role in the evolution and activity of galaxies where the stellar formation process rejuvenates the population of massive stars (clearly this is not true in elliptical galaxies which harbour old stellar populations devoid of massive stars).

The feedback of massive stars on their environment is now widely recognised as an important process by the astrophysical community. This paper emphasizes the specific influence of massive stars on processes relevant to interstellar chemistry. In particular we examine how massive stars provide the necessary 'ingredients' for various physico-chemical processes.

\section{The influence of massive stars}\label{influence}
\begin{figure*}
\begin{center}
\includegraphics[width=0.9\textwidth]{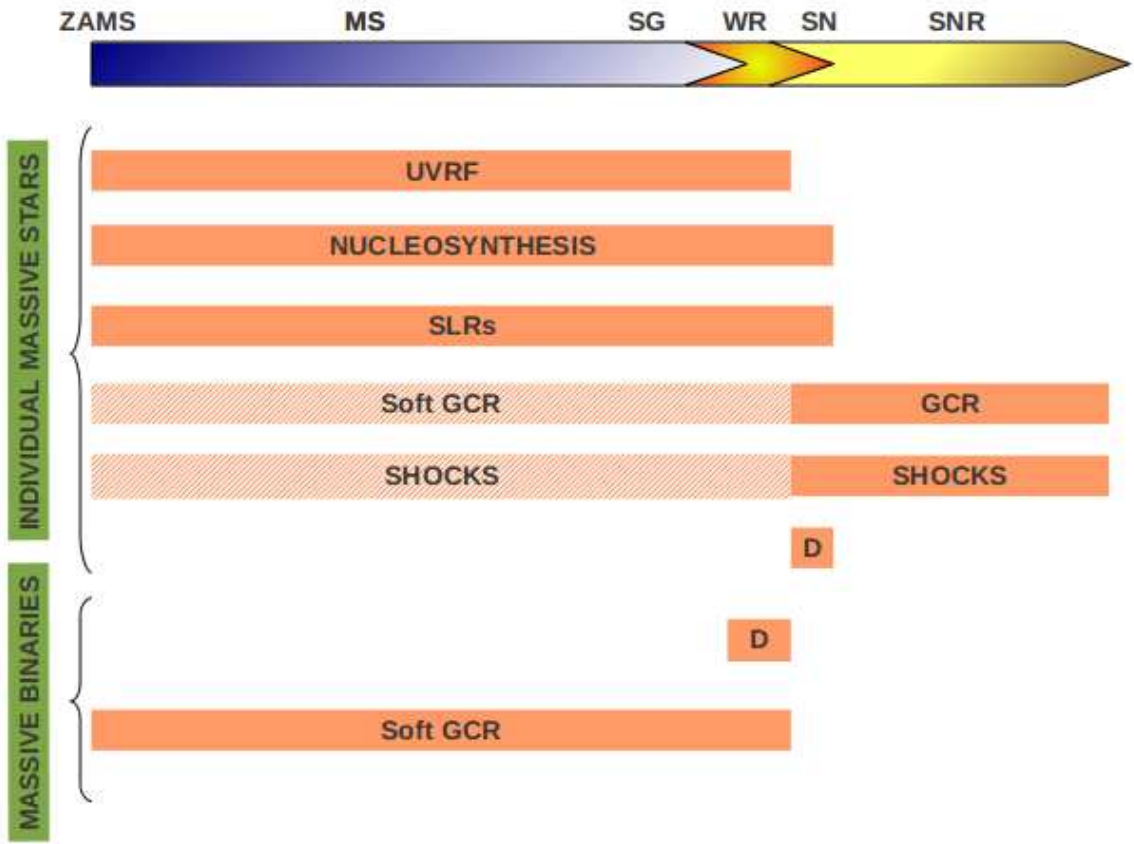}
\caption{Synthesis of the various aspects of massive star physics influencing astrochemical processes. The upper banner illustrates the evolutionary stages of massive stars, starting with the Zero Age Main-Sequence (ZAMS) and ending with the Supernova Remnant (SNR). The intermediate stages are, respectively, the main-sequence (MS), the supergiant phase (SG), the Wolf-Rayet (WR) stage, and the supernova (SN) explosion. Processes related to individual stars are summarized in the upper part, and processes specific to the case of colliding-wind massive binaries are illustrated in the lower part. The position of each horizontal bar indicates the evolutionary stages during which the associated process acts. Among these processes, UVRF stands for the action of ultraviolet radiation field, GCR for Galactic Cosmic-Rays, SLR for short-lived radionuclides, and D accounts for dust production.}
\label{summary}
\end{center}
\end{figure*}

A synthesis of the physical processes playing a major role in astrochemistry and which are related to massive stars -- either individual or in binaries -- is given in Figure\,\ref{summary}. Acronyms and the related processes are individually defined and discussed below. Some aspects of massive stars physics may occur only during particular evolutionary stages, or only in binary (or higher multiplicity) systems of massive stars. Note that processes associated to single stars are also at work in individual components of massive binaries. 

\subsection{Chemical enrichment}
An initial requirement for molecular formation is the availability of chemical elements. Beside the light elements that arose from the cosmological nucleosynthesis \citep{BBN}, several elements are produced through nuclear reactions in stellar cores. As far as nuclear fusion is concerned, massive stars are the only stellar objects able to synthesize nuclides heavier than carbon and oxygen, up to iron and nickel \citep{rauscher2001}. Through their strong stellar winds, large amounts of these elements are expelled into the interstellar medium during the evolution of a massive star \citep{dray2003,nomoto2013}. In particular, C- and O-enriched WR stars (WC and WO classes, respectively) feed the surrounding interstellar medium with large amounts of C and O. As massive stars evolve on time scales of a few Myr, they generally do not have the time to leave their formation site before exploding, and the chemical enrichment takes place close to their parent molecular cloud along the spiral arms of the Galaxy. During the supernova explosion, rapid neutron capture \citep{B2FH} reactions contribute further to fill the chart of nuclides (and the Mendeleiev table),  and chemically enriched material is expelled in large quantities into the surrounding interstellar medium. As a result, a wide range of chemical elements are made available to interstellar environments, where astrochemical processes take place.

\subsection{Radiation field}
Ultraviolet photons act as the most efficient destruction agent for small molecules through photodissociation processes \citep{tielens}. Any medium dominated by the ultraviolet radiation field (UVRF) of a massive star will therefore most probably be deficient in low molecular weight species. This constitutes a severe inhibition factor for the synthesis of large molecules, starting from small ones. The seminal work by \citet{tielenshollenbach1985II,tielenshollenbach1985I} presented astrochemical models of environments where ultraviolet photons are significant. Photoionization also plays a pivotal role in astrochemistry: it is the starting point of cationic chemical processes in transluscent interstellar clouds (i.e. clouds which are not dense enough to prevent UV photons permeating through them). For instance, in diffuse regions of the interstellar medium, carbon chemistry is initiated by the photoionization of carbon atoms, whose ionization energy is 11.26\,eV (thus, below 13.6\,eV, the ionization energy of the omnipresent atomic hydrogen). Ionic chemical pathways are of crucial importance in the interstellar medium \citep{ionchemism2008}, as they are characterized by rate constants up to several orders of magnitude higher than processes involving only neutral species \citep{umist}, resulting in a substantial decrease of reaction time scales. The action of UV light on chemical bondings is also responsible for the creation of free radicals, therefore activating interstellar chemistry via addition reactions without activation barriers, also characterized by higher rate constants than processes involving the breaking of chemical bonds. Such processes occur in gas phase, but also in interstellar ices containing small molecules, and they yield more complex chemical species \citep{iceuv2011}. The ultraviolet radiation field is also important in the formation process of fullerenes such as C$_{60}$ and C$_{70}$ \citep{formc602012}, recently identified in various astronomical environments \citep{fullercami,c60snellgren}.

Massive stars are the main providers of the ultraviolet radiation field in ``normal'' galaxies \citep{sternberg2003}. Because of their high effective temperatures, their continuum black body emission spectrum peaks in the far-ultraviolet domain. In addition, their brightness ($\propto$\,T$_\mathrm{eff}^4$) allows them to produce a wealth of photons in the energy range suitable to break chemical bonds, or to ionize neutral species. Ten thousand massive stars 10$^5$ times as bright as the Sun are as bright as one billion Suns, but with their emission spectrum peaking in the far-ultraviolet domain! Although the contribution of massive stars in number -- and mass -- in the global galactic stellar population is negligible, they definitely dominate it in brightness.

\subsection{Dust formation}
An important contribution to the content of the interstellar medium is the population of dust particles found, in particular, in dense clouds \citep{pagani2010}. Catalytic processes on dust grain surfaces play a pivotal role in astrochemistry \citep{grain1998,cazaux,tielens,h2o2dust2012}: the adsorption of chemical species onto grain surfaces substantially enhances the efficiency of chemical processes, allowing for the synthesis of many small and intermediate mass molecules whose formation would otherwise be severely inhibited in the gas phase. The most significant example is molecular hydrogen (H$_2$), but other abundant molecules such as water (H$_2$O) and formaldehyde (H$_2$CO) have increased concentrations too. The adsorption of atomic and molecular species onto dust grains induces a proximity between reaction partners that significantly enhances their probability to meet. Once the two partners meet on the surface, the probability to react is close to 1 unless a significant activation barrier inhibits the process. In addition, dust grains play the role of a 'third body' in chemical processes. They are indeed able to store the excess energy brought by reaction partners, therefore stabilizing the reaction products that would otherwise likely dissociate back to the reactants. For these reasons, dust particles turn out to be efficient interstellar chemical catalysts and without these grains, the molecular content of the interstellar medium would be substantially different.

Even though massive stars are not the main contributors to dust formation in the Universe \citep{gail2009,ventura2012}, their role should not be overlooked. In the 1970's, infrared observations revealed excess emission attributable to the presence of dust in the vicinity of some WR stars \citep{allen1972}. As evolved counterparts of O-type stars, WR stars display the signature of stellar nucleosynthesis, including nitrogen and helium enhancement in WN stars (due to the CNO process), and carbon enhancement with nitrogen depletion (due to the triple alpha process) in the case of the more evolved WC-type stars \citep{wrcrowther}. The latter objects are characterized by quite strong stellar winds, with mass loss rates up to 10$^{-5}$ solar mass per year. In addition to the WR stars catalogued by \citet{vdhcata}, more recent investigations in the infrared have revealed the existence of several tens of WR stars in the inner Galactic plane, including a large fraction of WC stars \citep{shara2009}. In massive binaries which harbour one of these WC stars, the powerful stellar winds interact, and the resulting interaction region reaches densities high enough to act as a nucleation site for dust particles \citep{williams1985,pwnwr104,marchenko1999,marchenkomoffat2007}. In such a hydrogen-poor environment, the precursors of these dust grains are likely to be small carbon clusters, or even silicon-bearing molecules such as silicon carbide: therefore dense WC stellar winds, surprisingly enough, are good candidates for the formation of molecular species \citep{cherchneff2000}. 

Supernovae are also known to eject large amounts of material including carbon and silicon. The subsequent nucleation (and condensation) of dust grains in the expanding shell -- starting about one year after the supernova explosion \citep{gehrz1990} -- provides an additional source of dust in the interstellar medium. Recent studies related to the nucleosynthesis of massive stars, point at a strong correlation between the composition of the ejected material from core-collapse supernovae and that of pre-solar carbon-rich dust grains \citep{ccsndust}, also in agreement with the idea that dust resulting from supernovae may be injected into distant proto-planetary disks \citep{ouellette2010}.
 
These contributions to the dust population are expected to play a role in astrochemical processes. The simultaneous statement that evolved massive stars produce dust, and that dust plays a significant role in astrochemistry through surface processes, lead to calls for dedicated theoretical and observational studies of chemical processes in dusty environments close to WC binaries or supernova remnants. 
New observatories such as the Atacama Large Millimeter Array in the Chilean Andes \citep{alma2004} will allow investigations with hitherto unprecedented sensitivity and angular resolution.

\subsection{High energy particles}
In dense molecular clouds that are opaque to UV light from neighboring stars, the main activation agent for chemical processes is the interaction of chemical species with Galactic Cosmic Rays (mainly charged H and He, along with other nuclides). These high energy particles interact with atoms and molecules and deposit a fraction of their energy along their path through the cloud. As a result, Cosmic Rays can dissociate and ionize \citep{cravens1978,crioniz2011} molecular species within dark clouds. Cosmic rays may even induce in situ the emission of UV photons \citep{PT-CR}.

Due to the low probability for  high energy particles to interact with atoms and molecules \citep{umist}, the rate constants for Cosmic-Ray induced processes are rather weak. Naively, one might therefore suppose that Galactic Cosmic Rays have relatively little influence on astrochemistry. However, two points must be made. First, since molecular clouds are big (at least several light-years across), the integrated effect due to the size of the cloud is significant: one should indeed remember that we are dealing with huge distances in such environments, involving large amounts of material despite the low densities. Second, if the cross sections were much larger, it would mean that high energy particles would lose a significant fraction of their energy before reaching the inner parts of such dark clouds. The consequence would be a stratification of dense molecular clouds into two main regions: a highly active outer layer influenced by interactions with Galactic Cosmic Rays, and a somewhat inert core where neither UV photons nor Cosmic Rays can permeate to deposit energy. In a Universe where Cosmic Ray cross sections would be high, the central parts of dense clouds -- constituting the cradle of stellar and planetary systems -- would be much less efficient chemical factories! The capability of Galactic Cosmic Rays to trigger chemical processes in interstellar ices is also significant. This is particularly true in collapsing dense clouds that are rich in various species, initially synthesized in the parent molecular cloud, and where the increasing densities (as a function of time, and with decreasing radial distance to the centre) favor the formation of icy mantles onto interstellar dust grains. Molecular species trapped in such ices may undergo Cosmic Ray induced dissociations, therefore creating free radicals likely to react with close molecules in the solid matrix \citep{pilling2011,pilling2012}. Such processes are very important for the molecular enrichment of interstellar environments on the way to the formation of planetary systems.

The role of massive stars in the production of Galactic Cosmic Rays is well established. Galactic Cosmic Rays are thought to be mainly accelerated by hydrodynamic shocks in supernova remnants (SNRs), resulting from the fast expansion of the outer shells of a massive star while the core shrinks to become a neutron star or a black hole \citep{ellison1997,prantzos2012,drury2012}. Supernova remnants are able to accelerate Cosmic Rays up to TeV energies \citep{drury2012,helderpasnr2012}. On the other hand, in binary systems of massive stars, the collision between stellar winds produces similar shocks but with a somewhat different geometry \citep[see e.g.][]{Pittard2009}. In such circumstances, the particle acceleration mechanism is also at work, so that massive binaries are able to accelerate particles as well \citep{EU,PD2006,DSACWB,debeckerreview}. A few tens of massive binaries are seen to accelerate particles -- the so-called particle-accelerating colliding-wind binaries (PACWBs) \citep{DW,farnieretacar,pacwbcata} -- but the maximum energy reached by particles in these objects should probably not reach levels above the TeV regime. Their contribution would therefore be limited to the production of soft Galactic Cosmic Rays, mainly in the MeV and GeV regimes. Even though the efficiency of supernova remnants to accelerate particles appears to be significantly higher than that of particle-accelerating colliding-wind binaries, the latter objects can accelerate particles over a period of several million years (compared to a few thousand years for supernova remnants). In addition, the high frequency (of the order of 50\,\%) of binaries among massive stars \citep{sanaevans2011} opens the possibility that particle acceleration may occur in many objects well before the supernova remnant phase. These facts suggest that the integrated contribution from the complete galactic population of particle-accelerating colliding-wind binaries is significant \citep{pacwbcata}. Further studies devoted to particle-accelerating colliding-wind binaries are needed -- and some are in progress -- to clarify their contribution to the production of soft Galactic Cosmic Rays, in order to estimate their influence on neighbouring molecular clouds populated by numerous molecules.

Finally, a further contribution to Galactic Cosmic Rays may arise from runaway massive stars. Runaway stars are stellar objects that have been ejected from their birth place, and are characterized by significantly large velocities with respect to the surrounding interstellar medium. This phenomenon can occur either via dynamic interactions in dense stellar clusters populated by many stars \citep{leonardduncan}, or following the supernova explosion in a binary system leading to the ejection of the companion star \citep{blaauw}. It is believed that about 10\,\% of massive stars could be runaways \citep{peri2012}. The important point here is that the strong stellar winds of these massive runaways interact with the interstellar medium, notably with molecular cloud material, therefore producing strong bow shocks. Such bow shocks are likely to be particle acceleration sites \citep{delvalleromero2012}, and should contribute to some extent to the in situ production of soft Galactic Cosmic Rays in the interstellar medium. Evidence for the particle acceleration activity of such objects has been provided by the detection of non-thermal radiation from such high energy particles, in the radio domain \citep{radiorunaway2010}, in X-rays \citep{aeaurigae2012}, and even in gamma-rays \citep{DRD2013}. All these examples provide strong evidence for the fact that massive stars are efficient providers of the high energy particles likely to intervene in astrochemical processes at work in molecular clouds.

\subsection{Shock formation}
In cold media such as dark molecular clouds which are characterized by temperatures of at most a few tens of K, processes such as neutral-neutral reactions with activation barriers are severely inhibited. The reaction rate constants of these processes are indeed very sensitive to the temperature. The existence of shocks in interstellar environments -- either due to massive star winds interacting with the surrounding medium, or following a supernova explosion as a massive star reaches its final evolutionary stage -- will significantly enhance chemical processes. Hydrodynamic shocks are efficient converters of kinetic energy into heat, and this will improve the kinetics of chemical processes that are sensitive to the local temperature \citep{umist}. For instance, shock-enhanced chemistry could play a role in the formation of polycyclic aromatic hydrocarbons (PAHs), starting from benzene. In this scenario, the energy required for cycloaddition is provided by the shock wave propagating through the interstellar medium \citep{pah-shock}. Shocks (that are potentially attributable to massive star winds) powering the Galactic Center region \citep{najarro1997} are also likely responsible for a significant heating of molecular clouds, as revealed recently by studies performed with the Herschel infrared satellite \citep{goicoechea2013}. The role played by shocks in the presence of ionic species such as CH$^+$ and SH$^+$ in various environments has also been pointed out by \citet{godard2012}.

Moreover, one should notice that the sputtering of dust grains -- and icy grain mantles -- that results from shock propagation in the interstellar medium will release large amounts of surface-processed molecules into the gas phase, therefore making these molecules available for further gas phase processing. Interstellar shocks also fragment dust grains. This process affects their size distribution, and can potentially cause their complete destruction \citep[see e.g.][]{JTH1996}.

Shock propagation in the interstellar medium as a result of supernova explosions can also compress molecular clouds to the point that a gravitational collapse is triggered, leading to the formation of stellar and planetary systems \citep{mcevol2012}. The resulting hot cores and protostellar objects constitute highly valuable astrochemical laboratories, where many molecules have already been identified \citep{vandishoeck2009}. Before the supernova explosion, the strong stellar winds of massive stars also compress the interstellar material. In places harbouring a dense population of massive stars, the combined effect of several stellar winds may produce large scale bubbles or superbubbles, providing further evidence of the ability of massive stars to shape and strongly influence the interstellar medium \citep{bubble1975}. As the stars evolve, the successive supernova explosions will further supply the mechanical energy required to enhance and expand the resulting superbubbles \citep{superbubble1981}. The boundaries of such bubbles contain hydrodynamic discontinuities that will influence the properties of interstellar regions of astrochemical interest.

\subsection{Radioactive nuclides}
Among short-lived radionuclides (SLRs),$^{26}$Al (half-life time, $\tau_{1/2}$ $\sim$ 0.72\,Myr) is synthesized by massive stars across all their evolutionary stages, from the main-sequence to the supernova \citep{palacios2005,limongi2006,gounellemeynet2012}. SLRs are injected into the interstellar medium through mass outflows, either slowly during the main-sequence or in a more spectacular way in supernova explosions. The best tracer of $^{26}$Al in the interstellar medium comes from the emission of $\gamma$-ray photons at 1.809\,MeV. Its distribution is closely cospatial with regions especially rich in massive stars \citep{prantzos1993,diehl2006}. In dense environments such as inner parts of protostellar disks, where stars and planetary systems form, the influence of $^{26}$Al becomes significant. The decay of $^{26}$Al leads to the production of excited $^{26}$Mg that relaxes radiatively through the release of $\gamma$-ray photons able to ionize chemical species \citep{markwick2002}. The presence of these SLRs can be considered as a pre-requisite for the activity of ionic chemical pathways in severely obscured protostellar environments. Another important short-lived radionuclide is $^{60}$Fe ($\tau_{1/2}$\,$\sim$\,1.5\,Myr). Unlike $^{26}$Al, its release occurs essentially during the supernova explosion. More recenly, \citet{cleeves2013} emphasized the role of SLRs in the ionization of proto-stellar disks, and computed ionization rates as a function of the location in the disk in the context of various disk models. 

The decay of short-lived radionuclides produced by massive stars therefore provides an in situ source of ionization, initiating ionic chemical pathways where neither interstellar nor proto-stellar ultraviolet radiation fields are present. Here again, the imprint of massive stars on astrochemical processes is identified, and demonstrates that these objects constitute important engines for the formation of molecules in various astronomical environments.

\section{Concluding remarks}\label{conclusions}
While massive stars are rare, their participation in physico-chemical processes in the interstellar medium and in proto-planetary environments is huge. They dominate the supply of chemical elements, they are the main providers of ionizing and dissociating photons, they accelerate high energy particles at various stages of their evolution, and they are responsible for a substantial input of mechanical energy into the interstellar medium. Hence massive stars are recognised to play a major role in the synthesis and transformation of molecular species in space. It is difficult to imagine how the 170 or so interstellar molecules identified so far could have been synthesized -- at least in presently measurable quantities -- in the absence of populations of massive stars disseminated along the spiral arms of the Milky Way. Indeed, the multiwavelength view of our galaxy provides compelling evidence that many tracers of material in various forms (dust and gas), or of various physical processes (for instance, release of short-lived radionuclides) match almost exactly, and display their highest signal in galactic regions where massive stars are the most probably located. In this context, the Galactic Center region is a key location to investigate interstellar chemistry: it indeed coincides with the highest abundances in many chemical elements heavier than helium, it is the densest part of the interstellar medium, and it is populated by a rich population of massive stars whose study is still far from complete. The importance of the Galactic Center in astrochemistry is demonstrated by the numerous studies that have been made of this region, including for instance recent results obtained with the Herschel infrared satellite \citep[e.g.][]{sonnentrucker2013,goicoechea2013,etxaluze2013}. In contrast, one may anticipate that elliptical galaxies contain a poorer astrochemistry; not only because of their low quantities of interstellar gas as compared to that of spiral galaxies, but also because of their lack of star formation activity likely to renew the population of massive stars. 

Clearly massive stars play an important role in affecting the molecular complexity of astrophysical environments. The exploration of astrochemistry therefore provides a highly valuable view of the feedback of massive stars on their environment, up to the highest levels of molecular complexity.

\acknowledgments
This is a pleasure to warmly thank Dr. Julian Mark Pittard for his kind and painstaking reading of the manuscript, that led to a substantial improvement of its quality. The author would also like to express his gratitude to the Swings family, father and son: Prof. Polydore Swings for his pioneering work on molecular spectroscopy applied to astrophysics in the 1930's and 1940's, and Prof. Jean-Pierre Swings for his careful reading and constructive comments on the present paper before submission, along with discussions related to his work that led to the first detection of infrared radiation from dust-making Wolf-Rayet stars in the 1970's. The Astrophysics Data System (ADS) database has been used for the bibliography.

%\nocite{*}
%\bibliographystyle{spr-mp-nameyear-cnd}
%\bibliography{myref}
%\bibliography{msac}

\end{document}